\documentclass{ws-ijmpa}
\usepackage[super,compress]{cite}
\usepackage{graphicx}
\usepackage{braket}
\usepackage{bm}
\usepackage{comment}
\usepackage{here}
\usepackage{multirow}
\usepackage{xcolor}

\begin{document}
\markboth{Ryota Minamizawa}{Magic square and three-zero textures for Dirac neutrinos}

\def\Journal#1#2#3#4{{#1} {\bf #2}, #3 (#4)}
\def\AHEP{Advances in High Energy Physics.} 
\def\ARNPS{Annu. Rev. Nucl. Part. Sci.} 
\def\AandA{Astron. Astrophys.} 
\def\ANP{Ann. Phys.}
\def\APJ{Astrophys. J.}
\def\APJS{Astrophys. J. Suppl}
\def\COMR{Comptes Rendues}
\def\CQG{Class. Quantum Grav.}
\def\CPC{Chin. Phys. C}
\def\EPJC{Eur. Phys. J. C}
\def\EPL{EPL}
\def\FSPAS{Front. Astron. Space Sci.}
\def\IJMPA{Int. J. Mod. Phys. A}
\def\IJMPE{Int. J. Mod. Phys. E}
\def\JCAP{J. Cosmol. Astropart. Phys.}
\def\JHEP{J. High Energy Phys.}
\def\JETPL{JETP. Lett.}
\def\JETPUSSR{JETP (USSR)}
\def\JPG{J. Phys. G} 
\def\JPCS{J. Phys. Conf. Ser.} 
\def\JPGNP{J. Phys. G: Nucl. Part. Phys.} 
\def\MPLA{Mod. Phys. Lett. A}
\def\NIMA{Nucl. Instrum. Meth. A.}
\def\NATU{Nature}
\def\NCA{Nuovo Cimento}
\def\NJP{New. J. Phys.}
\def\NPB{Nucl. Phys. B}
\def\NPBOLD{Nucl. Phys.}
\def\NPBSUPPL{Nucl. Phys. B. Proc. Suppl.}
\def\PL{Phys. Lett.}
\def\PLB{{Phys. Lett.} B}
\def\PMCA{PMC Phys. A}
\def\PREP{Phys. Rep.}
\def\PPNP{Prog. Part. Nucl. Phys.}
\def\PLBOLD{Phys. Lett.}
\def\PAN{Phys. Atom. Nucl.}
\def\PRL{Phys. Rev. Lett.}
\def\PRD{Phys. Rev. D}
\def\PRC{Phys. Rev. C}
\def\PR{Phys. Rev.}
\def\PTP{Prog. Theor. Phys.}
\def\PTEP{Prog. Theor. Exp. Phys.}
\def\RMP{Rev. Mod. Phys.}
\def\RPP{Rep. Prog. Phys.}
\def\SJNP{Sov. J. Nucl. Phys.}
\def\SCIENCE{Science}
\def\TNYAS{Trans. New York Acad. Sci.}
\def\ZETP{Zh. Eksp. Teor. Piz.}
\def\ZFPH{Z. fur Physik}
\def\ZPC{Z. Phys. C}

%
\catchline{}{}{}{}{}
%


\title{Magic square and three-zero textures for Dirac neutrinos}

\author{Ryota Minamizawa}

\address{Graduate School of Science, Tokai University,\\
4-1-1 Kitakaname, Hiratsuka, Kanagawa 259-1292, Japan}

\author{Yuta Hyodo}

\address{Graduate School of Science and Technology, Tokai University,\\
4-1-1 Kitakaname, Hiratsuka, Kanagawa 259-1292, Japan}

\author{Teruyuki Kitabayashi\footnote{Corresponding author}}

\address{Department of Physics, Tokai University,\\
4-1-1 Kitakaname, Hiratsuka, Kanagawa 259-1292, Japan\\
teruyuki@tokai-u.jp}

\maketitle

\begin{history}
\received{Day Month Year}
\revised{Day Month Year}
\end{history}

\begin{abstract}
We show a matrix decomposition of flavor mass matrix for Dirac neutrinos $M$ by sum as $M=M'+M^0$ where $M'$ obeys the feature of the magic square and $M^0$ is three-zero texture. The favorable three-zero textures in the context of magic square are explored. It turned out that so-called the normal ordering of neutrino masses is favored over the inverted ordering in the context of magic square.
\end{abstract}

\ccode{PACS numbers:14.60.Pq}


\section{Introduction\label{section:introduction}}
The texture of the flavor neutrino mass matrix is a long standing problem in particle physics. Many possible textures are proposed in the literature, such as, tri-bi maximal texture \cite{Harrison2002PLB,Xing2002PLB,Harrison2002PLB2,Kitabayashi2007PRD}, $\mu$-$\tau$ symmetric texture \cite{Fukuyama1997,Lam2001PLB,Ma2001PRL,Balaji2001PLB,Koide2002PRD,Kitabayashi2003PRD,Koide2004PRD,Aizawa2004PRD,Ghosal2004MPLA,Mohapatra2005PRD,Koide2005PLB,Kitabayashi2005PLB,Haba206PRD,Xing2006PLB,Ahn2006PRD,Joshipura2008EPJC,Gomez-Izquierdo2010PRD,He2001PRD,He2012PRD,Gomez-Izquierdo2017EPJC,Fukuyama2017PTEP,Kitabayashi2016PRD,Bao2021arXiv}, and textures under discrete symmetries e.g., $A_n$ and $S_n$ \cite{Altarelli2010PMP}.

One of the schemes of simple textures for neutrino mass matrix is known as texture zeros. \cite{Berger2001PRD,Frampton2002PLB,Xing2002PLB530,Xing2002PLB539,Kageyama2002PLB,Xing2004PRD,Grimus2004EPJC,Low2004PRD,Low2005PRD,Grimus2005JPG,Dev2007PRD,Xing2009PLB,Fritzsch2011JHEP,Kumar2011PRD,Dev2011PLB,Araki2012JHEP,Ludle2012NPB,Lashin2012PRD,Deepthi2012EPJC,Meloni2013NPB,Meloni2014PRD,Dev2014PRD,Felipe2014NPB,Ludl2014JHEP,Cebola2015PRD,Gautam2015PRD,Dev2015EPJC,Kitabayashi2016PRD1,Zhou2016CPC,Singh2016PTEP,Bora2017PRD,Barreiros2018PRD,Kitabayashi2018PRD,Barreiros2019JHEP,Capozzi2020PRD,Singh2020EPL,Barreiros2020,Kitabayashi2020PRD,Kitabayashi2017IJMPA,Kitabayashi2017IJMPA2,Kitabayashi2019IJMPA}. There are two usages of the term  ``texture zeros.'' To begin, if a flavor neutrino mass matrix contains zero elements, the texture of this flavor mass matrix is referred to as texture zeros. Second, even though the flavor neutrino mass matrix $M$ has no vanishing element, if a matrix $M$ is related to another matrix $M^0$ and $M^0$ has vanishing elements, we say that $M^0$ has a structure of the texture zeros for flavor neutrino mass matrix  $M$. For example, according to the seesaw mechanism for generating tiny neutrino masses, the flavor neutrino mass matrix is decomposed as $M \simeq M_{\rm D} M_{\rm R}^{-1} M_{\rm D}$, where $M_{\rm D}$ represents the Dirac mass matrix and $M_{\rm R}$ represents the mass matrix for heavy right-handed neutrinos. If the Dirac mass matrix has vanishing elements, e.g., $M_{\rm D} = M^0$, the texture of the Dirac mass matrix is called texture zeros. Aside from the terminology, the texture zeros scheme is concerned with the allowed number of zero elements in the mass matrix.  

In this paper, we use the term ``texture zeros'' as a second meaning. Instead, of the usual matrix decomposition by-product such as $M \simeq M^0 M_{\rm R}^{-1} M^0$, we take a matrix decomposition by sums such as $M=M'+M^0$ where $M'$ does not have zero elements and $M^0$ has zero elements. We show that if the matrix $M'$ obeys the magic square's feature \cite{Levitin2011,Sallows1997MathIntelli,Loly2009LLA,Nordgren2012LAA,Nordgren2020,Nordgren2021,Borsten2017CQG}, the texture of $M^0$ can naturally become a three-zero texture.

We note that the structure of the neutrino flavor mass matrix under symmetries with magic square has been studied previously \cite{Harrison2004PLB,Lam2006PLB,Gautam2016PRD,Yang2021arXiv,Channey2019JPGNP,Verma2020JPGNP,Hyodo2020IJMPA,Hyodo2021PTEP}, however, the constraint of the number of zero elements in the flavor neutrino mass matrix with magic square under decomposition by sum $M=M'+M^0$ is revealed for the first time in this paper. 

This paper is organized as follows. In section \ref{section:decomposition}, we present a brief review the of magic square and make a matrix decomposition of flavor mass matrix for Dirac neutrinos by sum. Then, we show that if the matrix $M'$ obeys the property of the magic square, the texture of $M^0$ can naturally become three-zero texture. In addition,  24 three-zero textures that can be compared with observations are shown concretely. Section \ref{section:numerical}, selects the most favorable three-zero texture based on numerical calculations. A summary is provided in Section \ref{section:summary}.

\section{Magic square and three-zero textures \label{section:decomposition}}
\subsection{Magic square}
A magic square of order $n$ is a $n \times n$ square grid filled with distinct natural numbers \cite{Levitin2011,Sallows1997MathIntelli,Loly2009LLA,Nordgren2012LAA,Nordgren2020,Nordgren2021,Borsten2017CQG}. Each cell contains a number in the range $1,2,\cdots,n^2$. The total of the numbers in each row, column, and diagonal is equal. For example, a magic square of order 3 is schematically shown as
\begin{eqnarray}
\begin{array}{|c|c|c|}
\hline
 \ 2 \ & \  7 \  & \ 6 \ \\
\hline
 9 & 5 & 1  \\
\hline
 4 & 3 & 8 \\  
\hline
\end{array}
\begin{array}{c}
\leftarrow 15 \\
\leftarrow 15 \\
\leftarrow 15 \\
\end{array}
\nonumber \\
\begin{array}{ccccccc}
& \nearrow & ~\uparrow & ~\uparrow & ~\uparrow & \nwarrow &\\
15 & & ~15 & ~ 15 & ~15 & &15  \\
\end{array} 
\end{eqnarray}
where the sum (which is called magic constant or magic sum) is 15. 

It is known that the following Lucas's matrix is a simple realization of the magic squares\cite{Sallows1997MathIntelli,Loly2009LLA,Nordgren2012LAA,Nordgren2020,Nordgren2021}: 
\begin{eqnarray}
M^{\rm magic}
= \left(
\begin{matrix}
\gamma-\beta & \gamma+\alpha+\beta  & \gamma-\alpha  \\
\gamma-\alpha+\beta  & \gamma  & \gamma+\alpha-\beta  \\
\gamma+\alpha  & \gamma-\alpha-\beta  & \gamma+\beta 
\end{matrix}
\right),
\label{Eq:Mmagic}
\end{eqnarray}
where, $\alpha, \beta$ and $\gamma$ are complex numbers.

\subsection{Decomposition and three-zero textures}
With Lucas's matrix, the Dirac flavor neutrino mass matrix
\begin{eqnarray}
M=
\left(
\begin{matrix}
a & b & c  \\
d  & e  & f  \\
g  & h  & i 
\end{matrix}
\right),
\end{eqnarray}
can be decomposed
\begin{eqnarray}
M=M^{\rm magic} + M^{\rm 0} ,
\label{Eq:M=Mmagic+deltaM}
\end{eqnarray}
where
\begin{eqnarray}
 M^{\rm 0} 
= \left(
\begin{matrix}
a_0 & b_0  & c_0  \\
d_0  & e_0  & f_0  \\
g_0  & h_0  & i_0 
\end{matrix}
\right).
\label{Eq:deltaM}
\end{eqnarray}
This decomposition applies to any arbitrary complex matrix, not just the Dirac flavor neutrino mass matrix.

The number of complex elements in the complex matrix $M$ is nine. Since the number of complex elements in Lucas's form $M^{\rm magic}$ is three, the number of independent complex elements in $M^{\rm 0} $ is six. Three complex elements in $M^{\rm 0} $ can vanish without loss of generality. As a result if we decompose the Dirac flavor neutrino mass matrix by sum with  $M^{\rm magic}$ and $M^{\rm 0} $ as shown in Eq.(\ref{Eq:M=Mmagic+deltaM}), the most natural form of $M^{\rm 0} $ may be three-zero textures. 

\subsection{Three-zero textures}
We construct concrete three-zero textures $M_0$ for comparison with observations.
The nine elements of Lucas's matrix are a linear combination of $\alpha,\beta,\gamma$. The (2,2) component of Lucas's matrix is composed simply of one complex element $\gamma$. \cite{Sallows1997MathIntelli} As a result of the experimental results, we can directly identify $\gamma$. We take $e = \gamma$, in other words $e_0=0$.  In this case, the following 28 three-zero textures are obtained:
\begin{eqnarray}
M_1^{\rm 0} 
= \left(
\begin{matrix}
0 & 0  &c_0  \\
d_0  & 0  &f_0  \\
g_0  &h_0  &i_0 
\end{matrix}
\right)
, \quad
M_2^{\rm 0} 
= \left(
\begin{matrix}
0 &b_0  & 0  \\
d_0  & 0  &f_0  \\
g_0  &h_0  &i_0 
\end{matrix}
\right)
, \quad
M_3^{\rm 0} 
= \left(
\begin{matrix}
0 &b_0  &c_0  \\
0  & 0  &f_0  \\
g_0  &h_0  &i_0 
\end{matrix}
\right), \nonumber 
\end{eqnarray}
\begin{eqnarray}
M_4^{\rm 0} 
= \left(
\begin{matrix}
0 &b_0  &c_0  \\
d_0  & 0  & 0  \\
g_0  &h_0  &i_0 
\end{matrix}
\right)
, \quad
M_5^{\rm 0} 
= \left(
\begin{matrix}
0 &b_0  &c_0  \\
d_0  & 0  &f_0  \\
0  &h_0  &i_0 
\end{matrix}
\right)
, \quad
M_6^{\rm 0} 
= \left(
\begin{matrix}
0 &b_0  &c_0  \\
d_0 & 0  &f_0  \\
g_0  & 0  &i_0 
\end{matrix}
\right), \nonumber
\end{eqnarray}
\begin{eqnarray}
M_7^{\rm 0} 
= \left(
\begin{matrix}
a_0 & 0  & 0  \\
d_0  & 0  &f_0  \\
g_0  &h_0  &i_0 
\end{matrix}
\right)
, \quad
M_8^{\rm 0} 
= \left(
\begin{matrix}
a_0 & 0  &c_0  \\
0  & 0  &f_0  \\
g_0 &h_0  &i_0 
\end{matrix}
\right),
\quad
M_9^{\rm 0} 
= \left(
\begin{matrix}
a_0 & 0  &c_0  \\
d_0  & 0  & 0  \\
g_0 &h_0  &i_0 
\end{matrix}
\right),
\nonumber
\end{eqnarray}
\begin{eqnarray}
M_{10}^{\rm 0} 
= \left(
\begin{matrix}
a_0& 0  &c_0  \\
d_0  & 0  &f_0  \\
0  &h_0  &i_0 
\end{matrix}
\right)
, \quad
M_{11}^{\rm 0} 
= \left(
\begin{matrix}
a_0& 0  &c_0  \\
d_0 & 0  &f_0  \\
g_0  &h_0  & 0 
\end{matrix}
\right)
, \quad
M_{12}^{\rm 0} 
= \left(
\begin{matrix}
a_0 &b_0  & 0  \\
0  & 0  &f_0  \\
g_0  &h_0  &i_0 
\end{matrix}
\right),
\nonumber
\end{eqnarray}
\begin{eqnarray}
M_{13}^{\rm 0} 
= \left(
\begin{matrix}
a_0 &b_0  & 0  \\
d_0  & 0  & 0  \\
g_0 &h_0  &i_0 
\end{matrix}
\right)
, \quad
M_{14}^{\rm 0} 
= \left(
\begin{matrix}
a_0&b_0  & 0  \\
d_0  & 0  &f_0  \\
g_0  & 0  &i_0 
\end{matrix}
\right)
, \quad
M_{15}^{\rm 0} 
= \left(
\begin{matrix}
a_0&b_0  & 0  \\
d_0  & 0  &f_0  \\
g_0  &h_0  & 0 
\end{matrix}
\right), 
 \nonumber
\end{eqnarray}
\begin{eqnarray}
M_{16}^{\rm 0} 
= \left(
\begin{matrix}
a_0 &b_0  &c_0  \\
0  & 0  &f_0  \\
0  &h_0  &i_0 
\end{matrix}
\right)
, \quad
M_{17}^{\rm 0} 
= \left(
\begin{matrix}
a_0 &b_0  &c_0  \\
0  & 0  &f_0  \\
g_0  & 0  &i_0 
\end{matrix}
\right)
, \quad
M_{18}^{\rm 0} 
= \left(
\begin{matrix}
a_0 &b_0  &c_0  \\
0  & 0  &f_0  \\
g_0  &h_0  & 0 
\end{matrix}
\right),
 \nonumber
\end{eqnarray}
\begin{eqnarray}
M_{19}^{\rm 0} 
= \left(
\begin{matrix}
a_0 &b_0  &c_0  \\
d_0  & 0  & 0  \\
0  &h_0  &i_0 
\end{matrix}
\right)
, \quad
M_{20}^{\rm 0} 
= \left(
\begin{matrix}
a_0 &b_0  &c_0  \\
d_0  & 0  & 0  \\
g_0  & 0  &i_0 
\end{matrix}
\right)
, \quad
M_{21}^{\rm 0} 
= \left(
\begin{matrix}
a_0 &b_0  &c_0  \\
d_0  & 0  & 0  \\
g_0  &h_0  & 0 
\end{matrix}
\right),
 \nonumber
\end{eqnarray}
\begin{eqnarray}
M_{22}^{\rm 0} 
= \left(
\begin{matrix}
a_0 &b_0  &c_0  \\
g_0  & 0  &f_0  \\
0  & 0  &i_0 
\end{matrix}
\right)
, \quad
M_{23}^{\rm 0} 
= \left(
\begin{matrix}
a_0 &b_0  &c_0  \\
d_0  & 0  &f_0  \\
0  &h_0  & 0 
\end{matrix}
\right)
, \quad
M_{24}^{\rm 0} 
= \left(
\begin{matrix}
a_0 &b_0  &c_0  \\
d_0  & 0  &f_0  \\
g_0 & 0  & 0 
\end{matrix}
\right),
\label{Eq:deltaMi}
\end{eqnarray}
and
\begin{eqnarray}
M_{25}^{\rm 0} 
&=&\left(
\begin{matrix}
0 &b_0  &c_0  \\
d_0  & 0  &f_0  \\
g_0  &h_0  & 0 
\end{matrix}
\right),
\quad
M_{26}^{\rm 0} 
=
\left(
\begin{matrix}
a_0 & 0  &c_0  \\
d_0 & 0  &f_0  \\
g_0 & 0  &i_0 
\end{matrix}
\right),
\quad
M_{27}^{\rm 0} 
=
\left(
\begin{matrix}
a_0 &b_0  & 0  \\
d_0 & 0  &f_0  \\
0  &h_0  &i_0 
\end{matrix}
\right), 
\nonumber \\
M_{28}^{\rm 0} 
&=&
\left(
\begin{matrix}
a_0 &b_0  &c_0  \\
0  & 0  & 0  \\
g_0  &h_0  &i_0 
\end{matrix}
\right).
\label{Eq:deltaMi_omitted}
\end{eqnarray}

For 24 three-zero textures of $M_1^{\rm 0} , M_2^{\rm 0} , \cdots , M_{24}^{\rm 0}$ in Eq.(\ref{Eq:deltaMi}), the Dirac flavor neutrino mass matrix $M$ is obtained as
\begin{eqnarray}
M= M_i^{\rm magic} + M_i^{\rm 0} , \quad (i=1,2,\cdots,24),
\end{eqnarray}
 and we can determine the non-zero elements in $M_i^{\rm 0} $ and $M_i^{\rm magic}$ in the terms of the elements $a, b, c,\cdots , i$ in the Dirac flavor neutrino mass matrix as follows:
\begin{eqnarray}
M_1^{\rm magic} : && \alpha=a+b-2e, \quad \beta=-a+e, \quad \gamma=e, \\
M_1^{\rm 0} : &&c_0 = a+b+c-3e, \quad d_0 =2a+b+d-4e, \quad f_0 =-2a-b+2e+f, \nonumber \\
&&g_0 =-a-b+e+g, \quad h_0 =b-2e+h, \quad i_0 = a-2e+i,  \nonumber
\end{eqnarray}
\begin{eqnarray}
M_2^{\rm magic} : && \alpha= -c + e, \quad \beta=-a + e, \quad \gamma=e, \\
M_2^{\rm 0} : &&b_0 = a + b + c - 3 e, \quad d_0 = a - c + d - e, \quad f_0 = -a + c - e + f, \nonumber \\
&&g_0 =c - 2 e + g, \quad h_0 = -a - c + e + h, \quad i_0 = a - 2 e + i, \nonumber 
\end{eqnarray}
\begin{eqnarray}
M_3^{\rm magic} : && \alpha=-a - d + 2 e, \quad \beta=-a + e, \quad \gamma=e, \\
M_3^{\rm 0} : &&b_0 = 2 a + b + d - 4 e, \quad c_0 = -a + c - d + e, \quad f_0 = d - 2 e + f, \nonumber \\
&&g_0 = a + d - 3 e + g, \quad h_0 = -2 a - d + 2 e + h, \quad i_0 = a - 2 e + i, \nonumber  
\end{eqnarray}
\begin{eqnarray}
M_4^{\rm magic}: && \alpha=-a + f, \quad \beta=-a + e, \quad \gamma=e, \\
M_4^{\rm 0} : &&b_0 = 2 a + b - 2 e - f, \quad c_0 = -a + c - e + f, \quad d_0 = d - 2 e + f, \nonumber \\
&&g_0 = a - e - f + g, \quad h_0 = -2 a + f + h, \quad i_0 = a - 2 e + i, \nonumber 
\end{eqnarray}
\begin{eqnarray}
M_5^{\rm magic} : && \alpha=-e + g, \quad \beta=-a + e, \quad \gamma=e, \\
M_5^{\rm 0} : &&b_0 = a + b - e - g, \quad c_0 = c - 2 e + g, \quad d_0 = a + d - 3 e + g, \nonumber \\
&&f_0 = -a + e + f - g, \quad h_0 = -a - e + g + h, \quad i_0 = a - 2 e + i, \nonumber 
\end{eqnarray}
\begin{eqnarray}
M_6^{\rm magic} : && \alpha=a - h, \quad \beta=-a + e, \quad \gamma=e, \\
M_6^{\rm 0} : &&b_0 = b - 2 e + h, \quad c_0 = a + c - e - h, \quad d_0 = 2 a + d - 2 e - h, \nonumber \\
&&f_0 = -2 a + f + h, \quad g_0 = -a - e + g + h, \quad i_0 = a - 2 e + i, \nonumber  
\end{eqnarray}
\begin{eqnarray}
M_7^{\rm magic}: && \alpha=-c + e, \quad \beta=b + c - 2 e, \quad \gamma=e, \\
M_7^{\rm 0} : &&a_0 = a + b + c - 3 e, \quad d_0 = -b - 2 c + d + 2 e, \quad f_0 = b + 2 c - 4 e + f, \nonumber \\
&&g_0 = c - 2 e + g, \quad h_0 = b - 2 e + h, \quad i_0 = -b - c + e + i, \nonumber 
\end{eqnarray}
\begin{eqnarray}
M_8^{\rm magic} : && \alpha=\frac{1}{2}(b-d), \quad \beta=\frac{1}{2}(b+d-2e), \quad \gamma=e, \\
M_8^{\rm 0} : &&a_0 = \frac{1}{2}(2 a + b + d - 4 e), \quad c_0 = \frac{1}{2}(b + 2 c - d - 2 e), \quad f_0 = d - 2 e + f, \nonumber \\
&&g_0 = \frac{1}{2}(-b + d - 2 e + 2 g), \quad h_0 = b - 2 e + h, \quad i_0 = \frac{1}{2}(-b - d + 2 i), \nonumber 
\end{eqnarray}
\begin{eqnarray}
M_9^{\rm magic} : && \alpha=\frac{1}{2}(b - 2 e + f), \quad \beta=\frac{1}{2}(b - f), \quad \gamma=e, \\
M_9^{\rm 0} : &&a_0 = \frac{1}{2}(2 a + b - 2 e - f), \quad c_0 = \frac{1}{2}(b + 2 c - 4 e + f), \quad d_0 = d - 2 e + f, \nonumber \\
&&g_0 = \frac{1}{2}(-b - f + 2 g), \quad h_0 = b - 2 e + h, \quad i_0 = \frac{1}{2}(-b - 2 e + f + 2 i), \nonumber
\end{eqnarray}
\begin{eqnarray}
M_{10}^{\rm magic} : && \alpha=-e + g, \quad \beta=b - g, \quad \gamma=e, \\
M_{10}^{\rm 0} : &&a_0 = a + b - e - g, \quad c_0 = c - 2 e + g, \quad d_0 = -b + d - 2 e + 2 g, \nonumber \\
&&f_0 = b + f - 2 g, \quad h_0 = b - 2 e + h, \quad i_0 = -b - e + g + i, \nonumber 
\end{eqnarray}
\begin{eqnarray}
M_{11}^{\rm magic} : && \alpha=b - i, \quad \beta=-e + i, \quad \gamma=e, \\
M_{11}^{\rm 0} : &&a_0 = a - 2 e + i, \quad c_0 = b + c - e - i, \quad d_0 = b + d - 2 i, \nonumber \\
&&f_0 = -b - 2 e + f + 2 i, \quad g_0 = -b - e + g + i, \quad h_0 = b - 2 e + h, \nonumber 
\end{eqnarray}
\begin{eqnarray}
M_{12}^{\rm magic} : && \alpha=-c + e, \quad \beta=-c + d, \quad \gamma=e, \\
M_{12}^{\rm 0} : &&a_0 = a - c + d - e, \quad b_0 = b + 2 c - d - 2 e, \quad f_0 = d - 2 e + f, \nonumber \\
&&g_0 = c - 2 e + g, \quad h_0 = -2 c + d + h, \quad i_0 = c - d - e + i, \nonumber \\\end{eqnarray}
\begin{eqnarray}
M_{13}^{\rm magic} : && \alpha=-c + e, \quad \beta=-c + 2 e - f, \quad \gamma=e, \\
M_{13}^{\rm 0} : &&a_0 = a - c + e - f, \quad b_0 = b + 2 c - 4 e + f, \quad d_0 = d - 2 e + f, \nonumber \\
&&g_0 = c - 2 e + g, \quad h_0 = -2 c + 2 e - f + h, \quad i_0 = c - 3 e + f + i, \nonumber 
\end{eqnarray}
\begin{eqnarray}
M_{14}^{\rm magic}c : && \alpha=-c + e, \quad \beta=c - h, \quad \gamma=e, \\
M_{14} ^{\rm 0} : &&a_0 = a + c - e - h, \quad b_0 = b - 2 e + h, \quad d_0 = -2 c + d + h, \nonumber \\
&&f_0 = 2 c - 2 e + f - h, \quad g_0 = c - 2 e + g, \quad i_0 = -c - e + h + i, \nonumber 
\end{eqnarray}
\begin{eqnarray}
M_{15}^{\rm magic} : && \alpha=-c + e, \quad \beta=-e + i, \quad \gamma=e, \\
M_{15}^{\rm 0} : &&a_0 = a - 2 e + i, \quad b_0 = b + c - e - i, \quad d_0 = -c + d + e - i, \nonumber \\
&&f_0 = c - 3 e + f + i, \quad g_0 = c - 2 e + g, \quad h_0 = -c - e + h + i, \nonumber 
\end{eqnarray}
\begin{eqnarray}
M_{16}^{\rm magic} : && \alpha=-e + g, \quad \beta=d - 2 e + g, \quad \gamma=e, \\
M_{16}^{\rm 0} : &&a_0 = a + d - 3 e + g, \quad b_0 = b - d + 2 e - 2 g, \quad c_0 = c - 2 e + g, \nonumber \\
&&f_0 = d - 2 e + f, \quad h_0 = d - 4 e + 2 g + h, \quad i_0 = -d + e - g + i, \nonumber 
\end{eqnarray}
\begin{eqnarray}
M_{17}^{\rm magic} : && \alpha=\frac{1}{2}(-d + 2 e - h), \quad \beta=\frac{1}{2}(d - h), \quad \gamma=e, \\
M_{17}^{\rm 0} : &&a_0 = \frac{1}{2}(2 a + d - 2 e - h), \quad b_0 = b - 2 e + h, \quad c_0 = \frac{1}{2}(2 c - d - h), \nonumber \\
&&f_0 = d - 2 e + f, \quad g_0 = \frac{1}{2}(d - 4 e + 2 g + h), \quad i_0 = \frac{1}{2}(-d - 2 e + h + 2 i), \nonumber 
\end{eqnarray}
\begin{eqnarray}
M_{18}^{\rm magic} : && \alpha=-d + i, \quad \beta=-e + i, \quad \gamma=e, \\
M_{18}^{\rm 0} : &&a_0 = a - 2 e + i, \quad b_0 = b + d - 2 i, \quad c_0 = c - d - e + i, \nonumber \\
&&f_0 = d - 2 e + f, \quad g_0 = d - e + g - i, \quad h_0 = -d - 2 e + h + 2 i, \nonumber
\end{eqnarray}
\begin{eqnarray}
M_{19}^{\rm magic} : && \alpha=-e + g, \quad \beta=-f + g, \quad \gamma=e, \\
M_{19}^{\rm 0} : &&a_0 = a - e - f + g, \quad b_0 = b + f - 2 g, \quad c_0 = c - 2 e + g, \nonumber \\
&&d_0 = d - 2 e + f, \quad h_0 = -2 e - f + 2 g + h, \quad i_0 = -e + f - g + i, \nonumber 
\end{eqnarray}
\begin{eqnarray}
M_{20}^{\rm magic} : && \alpha= \frac{1}{2}(f - h), \quad \beta=\frac{1}{2}(2 e - f - h), \quad \gamma=e, \\
M_{20}^{\rm 0} : &&a_0 = \frac{1}{2}(2 a - f - h), \quad b_0 = b - 2 e + h, \quad c_0 = \frac{1}{2}(2 c - 2 e + f - h), \nonumber \\
&&d_0 = d - 2 e + f, \quad g_0 = \frac{1}{2}(-2 e - f + 2 g + h), \quad i_0 = \frac{1}{2}(-4 e + f + h + 2 i), \nonumber 
\end{eqnarray}
\begin{eqnarray}
M_{21}^{\rm magic} : && \alpha=-2 e + f + i, \quad \beta=-e + i, \quad \gamma=e, \\
M_{21}^{\rm 0} : &&a_0 = a - 2 e + i, \quad b_0 = b + 2 e - f - 2 i, \quad c_0 = c - 3 e + f + i, \nonumber \\
&&d_0 = d - 2 e + f, \quad g_0 = e - f + g - i, \quad h_0 = -4 e + f + h + 2 i, \nonumber 
\end{eqnarray}
\begin{eqnarray}
M_{22}^{\rm magic} : && \alpha=-e + g, \quad \beta=2 e - g - h, \quad \gamma=e, \\
M_{22}^{\rm 0} : &&a_0 = a + e - g - h, \quad b_0 = b - 2 e + h, \quad c_0 = c - 2 e + g, \nonumber \\
&&d_0 = d - 4 e + 2 g + h, \quad f_0 = 2 e + f - 2 g - h, \quad i_0 = -3 e + g + h + i, \nonumber 
\end{eqnarray}
\begin{eqnarray}
M_{23}^{\rm magic} : && \alpha=-e + g, \quad \beta=-e + i, \quad \gamma=e, \\
M_{23}^{\rm 0} : &&a_0 = a - 2 e + i, \quad b_0 = b + e - g - i, \quad c_0 = c - 2 e + g, \nonumber \\
&&d_0 = d - e + g - i, \quad f_0 =-e + f - g + i, \quad h_0 = -3 e + g + h + i, \nonumber 
\end{eqnarray}
\begin{eqnarray}
M_{24}^{\rm magic} : && \alpha=2 e - h - i, \quad \beta=-e + i, \quad \gamma=e, \\
M_{24}^{\rm 0} : &&a_0 = a - 2 e + i, \quad b_0 = b - 2 e + h, \quad c_0 = c + e - h - i, \nonumber \\
&&d_0 = d + 2 e - h - 2 i, \quad f_0 = -4 e + f + h + 2 i, \quad g_0 = -3 e + g + h + i. \nonumber 
\end{eqnarray}

On the other hand, the rank of the remaining four matrices $M_{25}^{\rm 0},M_{26}^{\rm 0},M_{27}^{\rm 0},M_{28}^{\rm 0} $ in Eq.(\ref{Eq:deltaMi_omitted}) is not enough to determine non-zero elements $\{a_0,b_0, \cdots,i_0\}$ in $M^{\rm 0}$ and $\{\alpha,\beta,\gamma\}$ in $M^{\rm magic}$ in the term of the elements $\{a,b,c, \cdots,i\}$ in $M$.
In this paper, we pick up only the 24 three-zero textures $M_1^{\rm 0},M_2^{\rm 0},\cdots,M_{24}^{\rm 0} $.

\subsection{Usual magic symmetry and Magic square condition}
The usual magic symmetry is given by the $Z_2$ symmetry \cite{Yang2021arXiv}. In the following matrix
\begin{eqnarray}
M 
&=&\left(
\begin{matrix}
a+e & a+c+d  &a-b  \\
a-b+d  & a  &a+c+e  \\
a+c  &a-b+e  & a+d 
\end{matrix}
\right),
\label{Eq:UsualMagic}
\end{eqnarray}
sum of each row and column is equal to $3a-b+c+d+e$. This matrix $M$ is under the following $Z_2$ symmetry:
\begin{eqnarray}
S_2 M S_2^T=M,
\end{eqnarray}
where
\begin{eqnarray}
S_2 
&=&\left(
\begin{matrix}
1/3 & -2/3  &-2/3  \\
-2/3 & 1/3  &-2/3  \\
-2/3  &-2/3  & 1/3 
\end{matrix}
\right),
\end{eqnarray}
and
\begin{eqnarray}
S_2^2=&\left(
\begin{matrix}
1 & 0  &0  \\
0 & 1  &0  \\
0  &0  & 1 
\end{matrix}
\right),
\end{eqnarray}
which is usual magic symmetry \cite{Lam2006PLB}.

The matrix $M$ in Eq.(\ref{Eq:UsualMagic}) has 5 parameters, and the magic square matrix under discussion in this paper can be thought of as a further diagonal and anti-diagonal sum conditions added from here. The following two conditions may be used: 
\begin{enumerate}
\item ${\rm Tr}M=3a$,
\item anti-trace ${\rm aTr}M \equiv M_{13}+M_{22}+M_{31} = 3a$.
\end{enumerate}
From the first condition, we obtain $e = -d$. From the second condition, $ b = c$ is required. The matrix $M$ should be 
\begin{eqnarray}
M 
&=&\left(
\begin{matrix}
a-d & a+b+d  &a-b  \\
a-b+d  & a  &a+b-d  \\
a+b  &a-b-d  & a+d 
\end{matrix}
\right),
\label{Eq:MagicInThisPapar}
\end{eqnarray}
which is magic square matrix in this paper. 

Because the number of parameters is reduced from 5 to 3, the magic square condition in Eq.(\ref{Eq:MagicInThisPapar}) is extremely severe. The exact magic square condition should be violated for Dirac neutrinos and these violation will appear in three-zero textures. In the next subsection, we focus our attention on the three-zero textures in the context of the magic square.

\section{Favorable three-zero textures \label{section:numerical}}
\subsection{Criteria}
For previous section, we showed that 24 three-zero textures can be compared with observations. Which of 24 three-zero textures is the most favored?

In this paper, it is assumed that the Dirac flavor neutrino mass matrix $M$ can be decomposed by sum containing the Lucas's matrix $M^{\rm magic}$. This is the most important assumption. Thus, the case of $M$ = $M_i^{\rm magic}$ may be regarded as the complete favorable case. 

According to Ref.\cite{Xing2016RPP}, we use the following dimensionless parameters to quantify the strength of the violations of the magic-square feature
\begin{eqnarray}
\epsilon_i^{\ell m} = \left| \frac{M_{\ell m} - (M^{\rm magic}_i)_{\ell m}}{(M^{\rm magic}_i)_{\ell m}} \right|.
\label{Eq:epsilon_i}
\end{eqnarray}
In the most favorable case, we obtain  
\begin{eqnarray}
\epsilon_i = \frac{1}{6} \sum_{\ell,m=1,2,3} \epsilon_i^{\ell m} = 0.
\end{eqnarray}
where $1/6$ is a normalization factor.
We will estimate $\epsilon_i$ for neutrino parameters in $3 \sigma$ region and pick up its minimum values. If the minimum value $\epsilon_i^{\rm min}$ close to 0, we regard the combination $\{M_i^{\rm magic},M_i^{\rm 0} \}$ are favored with the context of the magic square.

\subsection{Neutrino parameters}
The flavor neutrino mass matrix for Dirac neutrinos is reconstructed as 
\begin{eqnarray}
M = U {\rm diag.}(m_1,m_2,m_3) V^\dag,
\label{Eq:DiracFlavorNeutrinoMassMatrix2}
\end{eqnarray}
with $m_1,m_2$ and $m_3$ being the neutrino mass eigenstates, $U$ being the left-handed lepton mixing matrix, and $V$ being the transformation matrix of three right-handed neutrinos \cite{Borgohain2021JPG}. We assume that the charged leptons mass matrix is diagonal and real. In this case, $U$ becomes the Pontecorvo-Maki-Nakagawa-Sakata mixing matrix \cite{Pontecorvo1957,Pontecorvo1958,Maki1962PTP,PDG}, and whose elements are 
\begin{eqnarray}
U_{e1} &=& c_{12}c_{13}, \quad U_{e 2} = s_{12}c_{13}, \quad U_{e 3} = s_{13} e^{-i\delta_{\rm CP}},  \nonumber \\
U_{\mu 1} &=&- s_{12}c_{23} - c_{12}s_{23}s_{13} e^{i\delta_{\rm CP}}, \nonumber \\
U_{\mu 2} &=&  c_{12}c_{23} - s_{12}s_{23}s_{13}e^{i\delta_{\rm CP}}, \quad U_{\mu 3} = s_{23}c_{13}, \nonumber \\
U_{\tau 1} &=& s_{12}s_{23} - c_{12}c_{23}s_{13}e^{i\delta_{\rm CP}}, \nonumber \\
U_{\tau 2} &=& - c_{12}s_{23} - s_{12}c_{23}s_{13}e^{i\delta_{\rm CP}}, \quad U_{\tau 3} = c_{23}c_{13}.
\end{eqnarray}
We used the abbreviations $c_{ij}=\cos\theta_{ij}$ and $s_{ij}=\sin\theta_{ij}$ where $\theta_{ij}$  ($ij$= 12, 23,13) is a mixing angle and the Dirac CP phase is denoted by $\delta_{\rm CP}$. 

The unitary matrix $V$ has three rotation angles $\phi_{ij}$ $(ij=12, 23,13)$, six phases $\omega_i$ ($i=1,2,\cdots,5$) and $\delta_V$ where $\phi_{ij} \in (0, \pi/2)$ and $\omega_i, \delta_V \in (-\pi, \pi)$. The matrix $V$ can be parametrize as \cite{Borgohain2021JPG,Senjanovic2016PRD} 
\begin{eqnarray}
V = {\rm diag.}(e^{i\omega_1}, e^{i\omega_2}, e^{i\omega_3}) \tilde V (\phi_{ij}, \delta_V) {\rm diag.}(e^{i\omega_4}, e^{i\omega_5}, 1) ,
\label{Eq:V}
\end{eqnarray}
where the elements of $\tilde V$ are written in a way similar to the PMNS matrix as
\begin{eqnarray}
\tilde V_{e1} &=& \tilde c_{12}\tilde c_{13}, \quad \tilde V_{e 2} = \tilde s_{12}\tilde c_{13}, \quad \tilde V_{e 3} = \tilde s_{13} e^{-i\delta_V},  \nonumber \\
\tilde V_{\mu 1} &=&- \tilde s_{12}\tilde c_{23} - \tilde c_{12}\tilde s_{23}\tilde s_{13} e^{i\delta_V}, \nonumber \\
\tilde V_{\mu 2} &=&  \tilde c_{12}\tilde c_{23} - \tilde s_{12}\tilde s_{23}\tilde s_{13}e^{i\delta_V}, \quad \tilde V_{\mu 3} = \tilde s_{23}\tilde c_{13}, \nonumber \\
\tilde V_{\tau 1} &=& \tilde s_{12}\tilde s_{23} - \tilde c_{12}\tilde c_{23}\tilde s_{13}e^{i\delta_V}, \nonumber \\
\tilde V_{\tau 2} &=& - \tilde c_{12}\tilde s_{23} - \tilde s_{12}\tilde c_{23}\tilde s_{13}e^{i\delta_V}, \quad \tilde V_{\tau 3} = \tilde c_{23}\tilde c_{13},
\end{eqnarray}
with the abbreviations $\tilde c_{ij}=\cos\phi_{ij}$ and $\tilde s_{ij}=\sin\phi_{ij}$.

Note that due to the unphysical phases corresponding to the rephasing of three left-handed neutrinos, the three phases $(\phi_e,\phi_\mu,\phi_\tau)$ appearing at the left side of the mixing matrix $U$. Therefore, we rewrite the Dirac flavor neutrino mass matrix as
\begin{eqnarray}
M = {\rm diag.}(e^{i\phi_e},e^{i\phi_\mu}, e^{i\phi_\tau}) U {\rm diag.}(m_1,m_2, m_3) V^\dag.
\label{Eq:DiracFlavorNeutrinoMassMatrix3}
\end{eqnarray}

A global analysis of current data shows the following the best fit values of the squared mass differences $\Delta m_{ij}^2=m_i^2-m_j^2$ and the mixing angles for the so-called normal mass ordering (NO), $m_1<m_2<m_3$, of the neutrinos  \cite{Esteban2020JHEP}:
\begin{eqnarray} 
\frac{\Delta m^2_{21}}{10^{-5} {\rm eV}^2} &=& 7.42^{+0.21}_{-0.20} \quad (6.82\rightarrow 8.04), \nonumber \\
\frac{\Delta m^2_{31}}{10^{-3}{\rm eV}^2} &=& 2.510^{+0.027}_{-0.027}\quad (2.430 \rightarrow 2.593), \nonumber \\
\theta_{12}/^\circ &=& 33.45^{+0.77}_{-0.75} \quad (31.27 \rightarrow 35.87), \nonumber \\
\theta_{23}/^\circ &=& 42.1^{+1.1}_{-0.9} \quad (39.7 \rightarrow 50.9), \nonumber \\
\theta_{13}/^\circ &=& 8.62^{+0.12}_{-0.12}\quad (8.25 \rightarrow 8.98), \nonumber \\
\delta_{\rm CP}/^\circ &=& 230^{+36}_{-25}\quad (144 \rightarrow 350),
\label{Eq:neutrino_observation_NO}
\end{eqnarray}
where the $\pm$ represents the $1 \sigma$ region and the parentheses denote the $3 \sigma$ region. On the other hand, for the so-called inverted mass ordering (IO), $m_3 < m_1 \lesssim m_2$, we have
\begin{eqnarray} 
\frac{\Delta m^2_{21}}{10^{-5} {\rm eV}^2} &=& 7.42^{+0.21}_{-0.20} \quad (6.82 \rightarrow 8.04), \nonumber \\
\frac{\Delta m^2_{32}}{10^{-3}{\rm eV}^2} &=& -2.490^{+0.026}_{-0.028}\quad (-2.574 \rightarrow -2.410), \nonumber \\
\theta_{12}/^\circ &=& 33.45^{+0.78}_{-0.75} \quad (31.27 \rightarrow 35.87), \nonumber \\
\theta_{23}/^\circ &=& 49.0^{+0.9}_{-1.3} \quad (39.8 \rightarrow 51.6), \nonumber \\
\theta_{13}/^\circ &=& 8.61^{+0.14}_{-0.12}\quad (8.24 \rightarrow 9.02), \nonumber \\
\delta_{\rm CP}/^\circ &=& 278^{+22}_{-30}\quad (194 \rightarrow 345).
\label{Eq:neutrino_observation_IO}
\end{eqnarray}
In addition, we have the constraint
\begin{eqnarray} 
\sum m_i < 0.12 ~{\rm eV},
\label{Eq:sumM}
\end{eqnarray}
from observation of the cosmic microwave background radiation \cite{Vagnozzi2017,Planck2018}.

\begin{table}[b]
\tbl{$\epsilon_i^{\rm min}$, and $\epsilon_i^{\ell m}$ when $\epsilon_i^{\rm min}$ is obtained. }
{\scalebox{0.85}{
{\begin{tabular}{|c|cc|ccccccccc|}
\hline
 & $M^0_i$ & $\epsilon_i^{\rm min}$ & $\epsilon_i^{11}$ & $\epsilon_i^{12}$ & $\epsilon_i^{13}$ & $\epsilon_i^{21}$ & $\epsilon_i^{22}$ & $\epsilon_i^{23}$ & $\epsilon_i^{31}$ & $\epsilon_i^{32}$ & $\epsilon_i^{33}$\\
\hline
\multirow{24}{*}{NO} & $M_{14}^0$ & 0.327&0.117&0.503&0.000&0.228&0.000&0.455&0.260&0.000&0.398 \\
& $M_{2}^0$ &0.337&0.000&0.522&0.000&0.244&0.000&0.527&0.260&0.143&0.326 \\
& $M_{17}^0$ &0.357&0.227&0.571&0.0768&0.000&0.000&0.547&0.170&0.000&0.552  \\
& $M_{11}^0$ & 0.358&0.633&0.000&0.215&0.468&0.000&0.386&0.302&0.146&0.000 \\
& $M_{6}^0$ & 0.367&0.000&0.503&0.262&0.277&0.000&0.596&0.240&0.000&0.326  \\
& $M_{12}^0$&0.368&0.252&0.559&0.000&0.000&0.000&0.547&0.172&0.125&0.554   \\
& $M_{24}^0$& 0.370&0.633&0.145&0.128&0.570&0.000&0.415&0.328&0.000&0.000 \\
& $M_{10}^0$ &0.376&0.263&0.000&0.639&0.434&0.000&0.127&0.000&0.445&0.348   \\
& $M_{7}^0$ & 0.381&0.556&0.000&0.000&0.657&0.000&0.0985&0.0690&0.350&0.556  \\
& $M_{4}^0$ & 0.387&0.000&0.700&0.332&0.497&0.000&0.000&0.264&0.315&0.216   \\
& $M_{3}^0$ & 0.389&0.000&0.443&0.406&0.000&0.000&0.406&0.420&0.452&0.208  \\
& $M_{1}^0$ &0.393	&0.000&0.000&0.683&0.267&0.000&0.331&0.293&0.610&0.176 \\
& $M_{5}^0$ &0.395	&0.000&0.721&0.447&0.329&0.000&0.220&0.000&0.436&0.216\\
& $M_{22}^0$ &0.399&0.375&0.590&0.385&0.267&0.000&0.319&0.000&0.000&0.455\\
& $M_{18}^0$ &0.402&0.146&0.209&0.761&0.000&0.000&0.354&0.400&0.543&0.000 \\
& $M_{9}^0$ &0.404&0.381&0.000&0.655&0.466&0.000&0.000&0.151&0.445&0.327\\
& $M_{15}^0$ &0.416&0.766&0.290&0.000&0.251&0.000&0.425&0.696&0.0688&0.000 \\
& $M_{20}^0$ &0.418&0.364&0.684&0.387&0.129&0.000&0.000&0.224&0.000&0.721 \\
& $M_{23}^0$ &0.421&0.115&0.541&0.382&0.322&0.000&1.132&0.000&0.0371&0.000 \\
& $M_{19}^0$ &0.425&0.496&0.266&0.639&0.466&0.000&0.000&0.000&0.368&0.317  \\
& $M_{8}^0$ &0.434&0.745&0.000&0.430&0.000&0.000&0.340&0.435&0.146&0.506\\
& $M_{16}^0$&0.437&0.417&0.106&0.178&0.000&0.000&0.686&0.000&0.261&0.976 \\
& $M_{21}^0$ &0.444&0.146&0.280&0.746&0.313&0.000&0.000&0.787&0.393&0.000\\
& $M_{13}^0$ &0.450&0.269&0.630&0.000&0.129&0.000&0.000&0.272&0.705&0.694  \\
\hline
\multirow{24}{*}{IO} &$M_{24}^0$&0.312&	0.103&0.0234&0.188&0.171&0.000&0.995&0.391&0.000&0.000\\
& $M_{11}^0$ &0.325&0.103&0.000&0.207&0.200&0.000&0.996&0.396&0.0517&0.000 \\
& $M_{6}^0$ &0.330&0.000&0.0234&0.108&0.141&0.000&0.989&0.359&0.000&0.357 \\
& $M_{15}^0$ &0.334&0.103&0.134&0.000&0.126&0.000&0.988&0.322&0.330&0.000\\
& $M_{1}^0$ &0.341&0.000&0.000&0.157&0.749&0.000&0.186&0.754&0.0931&0.109 \\
& $M_{2}^0$ &0.344	&0.000&0.0764&0.000&0.172&0.000&0.984&0.322&0.152&0.357 \\
& $M_{18}^0$&0.347&0.103&0.156&0.153&0.000&0.000&0.986&0.347&0.336&0.000 \\
& $M_{17}^0$&0.354&0.0906&0.0234&0.131&0.000&0.000&0.986&0.366&0.000&0.526 \\
& $M_{4}^0$ &0.366&0.000&0.0903&0.0682&0.780&0.000&0.000&0.922&0.194&0.141 \\
& $M_{3}^0$ &0.369&0.000&0.113&0.184&0.000&0.000&0.986&0.375&0.200&0.357 \\
& $M_{14}^0$ &0.372&0.161&0.350&0.000&1.007&0.000&0.0447&0.233&0.000&0.434 \\
& $M_{8}^0$ &0.374&0.109&0.000&0.128&0.000&0.000&0.986&0.367&0.0517&0.601 \\
& $M_{20}^0$ &0.374&0.140&0.350&0.0382&1.008&0.000&0.000&0.286&0.000&0.424 \\
& $M_{13}^0$ &0.380&0.123&0.405&0.000&1.008&0.000&0.000&0.233&0.0986&0.414 \\
& $M_{19}^0$ &0.385&0.0749&0.333&0.0525&1.189&0.000&0.000&0.000&0.222&0.439 \\
& $M_{7}^0$ &0.389&0.144&0.000&0.000&0.229&0.000&0.978&0.322&0.0517&0.612 \\
& $M_{22}^0$ &0.391&0.213&0.145&0.167&1.004&0.000&0.363&0.000&0.000&0.452 \\
& $M_{5}^0$ &0.391&0.000&0.332&0.0525&1.183&0.000&0.0543&0.000&0.272&0.454\\
& $M_{16}^0$ &0.394&0.506&0.133&0.133&0.000&0.000&0.203&0.000&1.349&0.0396 \\
& $M_{23}^0$ &0.396&0.0373&0.272&0.0834&	0.775&0.000&0.258&0.000&0.948&0.000 \\
& $M_{21}^0$ &0.405&0.225&0.290&0.151&0.588&0.000&0.000&0.843&0.334&0.000 \\
& $M_{12}^0$ &0.413&0.143&0.0948&0.000&0.000&0.000&1.054&0.287&0.769&0.130  \\
& $M_{9}^0$ &0.413	&0.178&0.000&0.519&0.774&0.000&0.000&0.281&0.607&0.121 \\
& $M_{10}^0$ &0.423&0.345&0.000&0.304&0.737&0.000&0.199&0.000&0.588&0.364 \\
\hline
\end{tabular}
}}
\label{tbl:epsilon_i_ell_m}}
\end{table}

\begin{table}[b]
\tbl{$\epsilon_i^{\rm min}$, and neutrino parameters when $\epsilon_i^{\rm min}$ is obtained. }
{\begin{tabular}{|c|cc|ccccccc|}
\hline
 & $M^0_i$ & $\epsilon_i^{\rm min}$ & $m_1$/eV & $m_2$/eV & $m_3$/eV & $\theta_{12}/^\circ$ & $\theta_{23}/^\circ$ & $\theta_{13}/^\circ$ & $\delta_{\rm CP}/^\circ$ \\
\hline
\multirow{24}{*}{NO} &$M_{14}^0$ & 0.327&0.0270&0.0283&0.0566&35.3&44.0&8.4&179.9 \\
& $M_{2}^0$ &0.337&0.0270&0.0283&0.0566&35.3&44.0&8.4&179.9 \\
& $M_{17}^0$ &0.357&0.0297&0.0310&0.0582&35.2&45.0&8.3&207.5  \\
& $M_{11}^0$ & 0.358&0.0294&0.0307&0.0582&34.5&48.1&8.4&223.8 \\
& $M_{6}^0$ & 0.367&0.0270&0.0283&0.0566&35.3&44.0&8.4&179.9  \\
& $M_{12}^0$&0.368&0.0297&0.0310&0.0582&35.2&45.0&8.3&207.5  \\
& $M_{24}^0$& 0.370&0.0294&0.0307&0.0582&34.5&48.1&8.4&223.8 \\
& $M_{10}^0$ &0.376&0.0221&0.0237&0.0541&32.5&48.2&8.8&219.6 \\
& $M_{7}^0$ & 0.381&0.0177&0.0196&0.0524&33.1&50.0&8.5	&146.7   \\
& $M_{4}^0$ & 0.387&0.0284&0.0296&0.0580&35.1&48.7&8.7	&277.4 \\
& $M_{3}^0$ & 0.389&0.0269&0.0282&0.0568&32.3&48.3&8.5&324.8  \\
& $M_{1}^0$ &0.393&0.0220&0.0237&0.0543&	31.9&49.8	&8.4&341.3 \\
& $M_{5}^0$ &0.395&0.0284&0.0296&0.0580&35.1&48.7&8.7&277.4\\
& $M_{22}^0$ &0.399&0.0266&0.0279&0.0566&33.0&44.5&8.4&184.9\\
& $M_{18}^0$ &0.402&0.0296&0.0309&0.0581&34.6&50.0&8.6&278.8 \\
& $M_{9}^0$ &0.404&0.0221&0.0237&0.0541&32.5&48.2&8.8&219.6\\
& $M_{15}^0$ &0.416&0.0298&0.0310&0.0589&34.7&42.3&8.9&327.1\\
& $M_{20}^0$ &0.418&0.0258&0.0273&0.0569&32.8&47.8&8.7&298.5 \\
& $M_{23}^0$ &0.421&0.0205&0.0221&0.0542&32.8&50.8&8.6&146.3 \\
& $M_{19}^0$ &0.425&0.0221&0.0237&0.0541&32.5&48.2&8.8&219.6  \\
& $M_{8}^0$ &0.434&0.0294&0.0307&0.0582&34.5&48.1&8.4&223.8\\
& $M_{16}^0$&0.437&0.0295&0.0308&0.0588&33.6&48.9&8.7&293.1 \\
& $M_{21}^0$ &0.444&0.0296&0.0309&0.0581&34.6&50.0&8.6&278.8\\
& $M_{13}^0$ &0.450&0.0258&0.0273&0.0569&32.8&47.8&8.7&298.5  \\
\hline
\multirow{24}{*}{IO} & $M_{24}^0$&0.312&0.0502&0.0510&0.00792&32.1&48.1&8.9&250.5\\
& $M_{11}^0$ &0.325&0.0502&0.0510&0.00792&32.1&48.1&8.9&250.5 \\
& $M_{6}^0$ &0.330&0.0502&0.0510&0.00792&32.1&48.1&8.9&250.5 \\
& $M_{15}^0$ &0.334&0.0502&0.0510&0.00792&32.1&48.1&8.9&250.5\\
& $M_{1}^0$ &0.341&0.0507&0.0514&0.0132&35.6&46.2&9.0&316.7 \\
& $M_{2}^0$ &0.344	&0.0502&0.0510&0.00792&32.1&48.1&8.9&250.5 \\
& $M_{18}^0$&0.347&0.0502&0.0510&0.00792&32.1&48.1&8.9&250.5 \\
& $M_{17}^0$&0.354&0.0502&0.0510&0.00792&32.1&48.1&8.9&250.5 \\
& $M_{4}^0$ &0.366&0.0512&0.0519&0.0116&	34.3&48.7&8.9&194.3 \\
& $M_{3}^0$ &0.369&0.0502&0.0510&0.00792&32.1&48.1&8.9&250.5 \\
& $M_{14}^0$ &0.372&0.0498&0.0505&0.00971&35.2&49.4&8.7&235.9 \\
& $M_{8}^0$ &0.374&0.0502&0.0510&0.00792&32.1&48.1&8.9&250.5 \\
& $M_{20}^0$ &0.374&0.0498&0.0505&0.00971&35.2&49.4&8.7&235.9 \\
& $M_{13}^0$ &0.380&0.0498&0.0505&0.00971&35.2&49.4&8.7&235.9 \\
& $M_{19}^0$ &0.385&0.0500&0.0508&0.00329&33.4&44.2&9.0&277.7 \\
& $M_{7}^0$ &0.389&0.0502&0.0510&0.00792&32.1&48.1&8.9&250.5 \\
& $M_{22}^0$ &0.391&0.0487&0.0496&0.00386&34.0&45.4&8.7&304.4 \\
& $M_{5}^0$ &0.391&0.0500&0.0508&0.00329&33.4&44.2&9.0&277.7\\
& $M_{16}^0$ &0.394&0.0514&0.0521&0.0129	&33.4&40.5&8.8&331.7 \\
& $M_{23}^0$ &0.396&0.0506&0.0513&0.0126&34.6&43.3&8.9&250.7 \\
& $M_{21}^0$ &0.405&0.0513&0.0520&0.0153&34.0&44.9&8.2&271.6 \\
& $M_{12}^0$ &0.413&0.0513&0.0519&0.0114&34.7&47.7&8.8&340.0  \\
& $M_{9}^0$ &0.413&0.0512&0.0519&0.0146&	32.0&47.8	&8.8&280.6 \\
& $M_{10}^0$ &0.423&0.0511&0.0519&0.0154&34.8&42.8&8.8&305.5 \\
\hline
\end{tabular}
\label{tbl:epsilon_i_neutrino_parameters}}
\end{table}

\subsection{Numerical analysis}
We randomly generated $10^7$ parameters sets $\{m_1,m_2,m_3,\theta_{12},\theta_{23},\theta_{13},\delta_{\rm CP}\}$ that satisfy the $3 \sigma$ region of Eq.(\ref{Eq:neutrino_observation_NO}), and Eq.(\ref{Eq:sumM}) for NO with $\phi_{ij} \in (0, \pi/2)$, $\omega_i, \delta_V, \phi_e, \phi_\mu, \phi_\tau \in (-\pi, \pi)$. We also randomly generated $10^7$ parameter sets that satisfy the $3 \sigma$ region of Eq.(\ref{Eq:neutrino_observation_IO}), and Eq.(\ref{Eq:sumM}), 
for IO. We estimate $\epsilon_i^{\rm min}$ using each parameter sets and the results are shown in Table.\ref{tbl:epsilon_i_ell_m} and Table.\ref{tbl:epsilon_i_neutrino_parameters}. Table.\ref{tbl:epsilon_i_ell_m} shows $\epsilon_i^{\rm min}$, and $\epsilon_i^{\ell m}$ when $\epsilon_i^{\rm min}$ is obtained. Table.\ref{tbl:epsilon_i_neutrino_parameters} shows $\epsilon_i^{\rm min}$, and neutrino parameters when $\epsilon_i^{\rm min}$ is obtained. 

First, we focus our attention on the NO case. Table.\ref{tbl:epsilon_i_ell_m} indicates that the $M_{14}^0$ ($\epsilon_{14}^{\rm min} = 0.327$) is the most favorable three-zero texture. The magnitude of $\epsilon_{14}^{\ell m}$ for non-zero element in $M_{14}^0$ is estimated as 
\begin{eqnarray}
&& \epsilon_{14}^{11}=0.117, \quad  \epsilon_{14}^{12}=0.503, \quad \epsilon_{14}^{21}=0.228,\nonumber \\
&& \epsilon_{14}^{23}=0.455, \quad \epsilon_{14}^{31}=0.0260, \quad  \epsilon_{14}^{33}=0.398.
\end{eqnarray}
Although the $(1,2)$ element violates the magic-square nature over $50\%$, the other elements obey the magic-square nature with $\epsilon_{14}^{\ell m} < 0.46$. More small $\epsilon_{14}^{\ell m}$ is preferable in the context of the magic square, however, we conclude that the decomposition of the Dirac flavor mass matrix with $M=M_{14}^{\rm magic} + M_{14}^0$ is reasonable and that the most favorable three-zero texture in the context of magic square is $M_{14}^0$ for NO.
 
The predicted neutrino parameters for the most favorable texture $M_{14}^0$ for NO are
\begin{eqnarray}
&& m_1 =  0.0270 {\rm eV}, \quad m_2= 0.0283 {\rm eV}, \quad m_3 = 0.0566 {\rm eV}, \nonumber \\
&& \theta_{12} = 35.3^\circ, \quad \theta_{23} = 44.0^\circ, \quad \theta_{13} = 8.4^\circ,\quad \delta_{\rm CP}=179.9^\circ,
\end{eqnarray}
as shown in Table. \ref{tbl:epsilon_i_neutrino_parameters}. We would like to comment on the T2K and NOvA tension\cite{T2K2021PRD,NOvA2021arXiv}. NOvA and T2K reported very different best fit value of $\delta_{\rm CP}$. In fact, for NO case, $\delta_{\rm CP} \sim 145^\circ $is reported from NOvA \cite{NOvA2021arXiv}, but it is reported as $\delta_{\rm CP} \sim 250^\circ$ from T2K \cite{T2K2021PRD}. Unfortunately, the predicted value $\delta_{\rm CP} \sim 179.9^\circ$ for the most favorable $M_{14}^0$ does not match the best fit values of $\delta_{\rm CP}$ from either NOvA or T2K; however, the predicted value is not excluded in the 90$\%$ C.L.\cite{NOvA2021arXiv,T2K2021PRD}.

%
\begin{figure}[t]
\begin{center}
\includegraphics[scale=0.78]{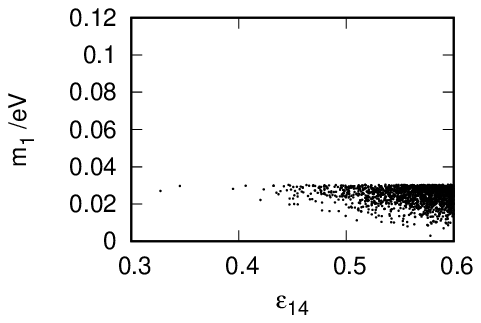}
\includegraphics[scale=0.78]{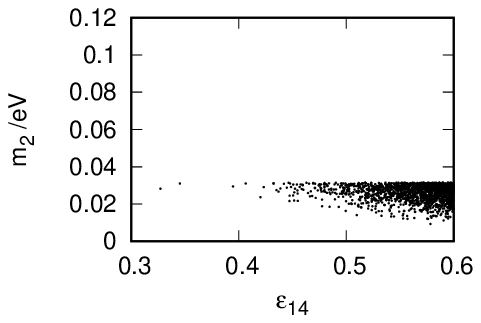}
\includegraphics[scale=0.78]{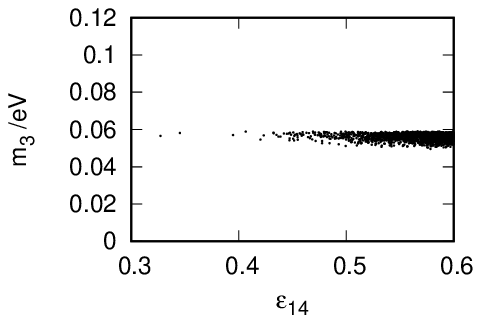}\\
\includegraphics[scale=0.78]{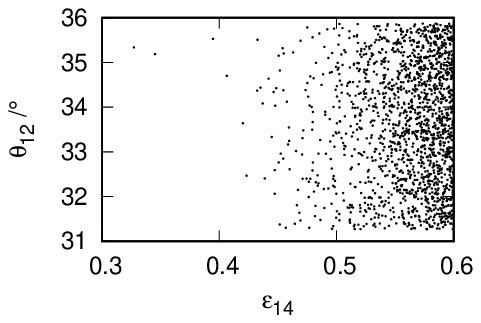}
\includegraphics[scale=0.78]{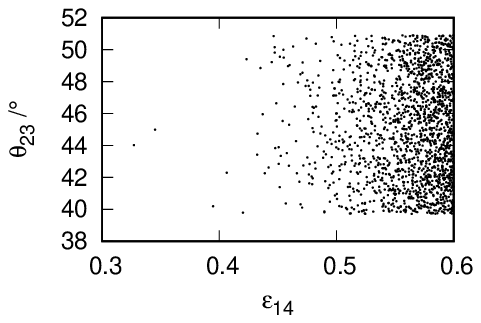}
\includegraphics[scale=0.78]{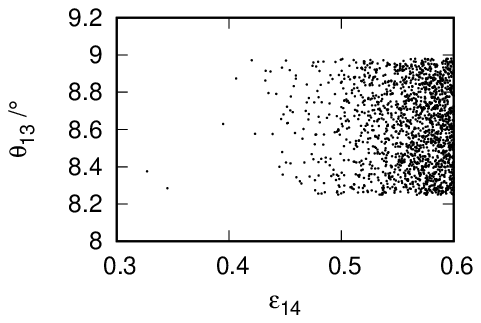}
\includegraphics[scale=0.78]{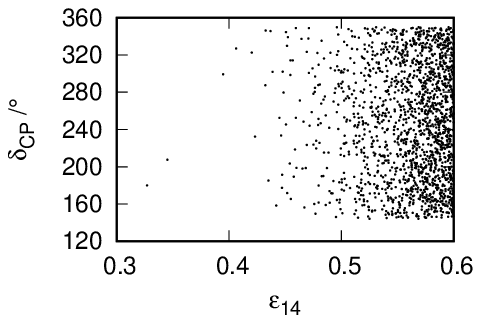}\\
\caption{Dependence of neutrino parameters $m_i,\theta_{ij} ,\delta_{\rm CP}$ on $\epsilon_i$ for $M^0_{14}$ in the case of NO.}
\label{fig:NO_14}
\end{center}
\end{figure}
%

Figure.\ref{fig:NO_14} shows the correlation between $\epsilon_{14}$ and neutrino parameters $m_1,m_2,m_3,\theta _{12},\theta_{23},\theta_{13},\delta_{\rm CP}$ for $M_{14}^0$ in the NO case. We have not plotted all $10^7$ points on Figure.\ref{fig:NO_14} to reduce the file size of the figure. Figure.\ref{fig:NO_14} indicates that there is no significant correlation between $\epsilon_{14}$ and neutrino parameters. Therefore, the small $\epsilon_{14}$ is realized under a fine-tuning of the neutrino parameters. Although this is undesirable result, we believe it is meaningful to show that the Dirac flavor mass matrix has a part of the nature of magic square for a specific parameters.

Next, we see the IO case. From Table.\ref{tbl:epsilon_i_ell_m}, the candidate of the favored three-zero textures in the context of a magic square is $M_{24}^0$ ($\epsilon_{24}^{\rm min} = 0.312$) and $M_{11}^0$ ($\epsilon_{11}^{\rm min} = 0.325$).  Although, these $\epsilon_{i}^{\rm min}$ are smaller than $\epsilon_{14}^{\rm min} = 0.327$ for NO, $\epsilon_{24}^{23}=0.995$ for $M_{24}^0$ and $\epsilon_{11}^{23}=0.996$ for $M_{11}^0$ violate the magic-square nature over $99\%$. Thus, we conclude that NO case is favored over IO in the context of magic square.

\section{Summary\label{section:summary}}
The number of zero elements permitted in the texture zeros neutrino flavor mass matrix is an important consideration. In this paper, we decomposed the Dirac flavor neutrino mass matrix as $M=M'+M^0$, where $M'$ is a matrix without zero elements and $M^0$ is a matrix with zero elements. We have shown that if $M'$ obeys a magic square, the three elements of $M^0$ can be naturally zero. The 24 three-zero textures that can be compared to the experimental results are numerically estimated, and we chose the preferred matrices in the context of magic square. It turned out that NO case is favored over IO, in the context of magic square.

\vspace{3mm}







\end{document}